\documentclass[12pt,epsf]{article}
\usepackage{amssymb,amsmath,amsbsy}
\usepackage{graphicx}
\usepackage{caption}
\usepackage{setspace}
\usepackage{subfigure}
\usepackage{cite} 
\newcommand{\be}{\begin{equation}}
\newcommand{\ee}{\end{equation}}
\newcommand{\bea}{\begin{eqnarray}}
\newcommand{\eea}{\end{eqnarray}}
\newcommand{\bear}{\begin{eqnarray}}
\newcommand{\eear}{\end{eqnarray}}
\newcommand{\beas}{\begin{eqnarray*}}

\newcommand{\eeas}{\end{eqnarray*}}
\newcommand{\ba}{\begin{array}}
\newcommand{\ea}{\end{array}}

\newcommand{\pd}[2][1]{\ifnum#1=1 \frac{\partial}{\partial {#2}} \else
  \frac{\partial^#1}{\partial {#2}^{#1}}\fi}
\newcommand{\dpd}[2][1]{\ifnum#1=1 \dfrac{\partial}{\partial {#2}} \else
  \frac{\partial^#1}{\partial {#2}^{#1}}\fi}
\newcommand{\td}[2][1]{\ifnum#1=1 \frac{d}{d{#2}} \else
  \frac{d^#1}{d{#2}^{#1}}\fi}

\renewcommand{\)}{\right)}

\newcommand{\nbox}{{\,\lower0.9pt\vbox{\hrule \hbox{\vrule height 0.2 cm \hskip 0.19 cm \vrule height 0.2 cm}\hrule}\,}}

\def\href#1#2{#2}

\usepackage{color}

\usepackage{xcolor}
\usepackage[citebordercolor=green, linkbordercolor={ 0 0 1}, linktocpage=true]{hyperref}

\begin{document}
\begin{titlepage}
\begin{NoHyper}
\hfill
\vbox{
    \halign{#\hfil         \cr
           } 
      }  
\vspace*{20mm}
\begin{center}
{\Large \bf  Reconstruction of an AdS Radiation/Boson Star Bulk Geometry Using Light-cone Cuts}

\vspace*{15mm}
\vspace*{1mm}
 Gabriel Trevi\~{n}o* 
\vspace*{.5cm}
\let\thefootnote\relax\footnote{*gabriel@physics.ucsb.edu}

{Department of Physics, University of California\\
Santa Barbara, CA 93106, USA}

\vspace*{1cm}
\end{center}
\begin{abstract}
Light-cone cuts have recently been proposed as a method to reconstruct the conformal metric of a holographic spacetime. We explore how additional information about the bulk geometry gets encoded in the structure of these light-cone cuts. In particular, we study how the hyperbolic angle related to a cusp in the light-cone cut encodes information about the matter content of the spacetime. We provide an explicit numerical example reconstructing the metric for a 4-dimensional spacetime composed by the superposition of a boson star and a gas of radiation in AdS.

\end{abstract}
\end{NoHyper}

\end{titlepage}
\vskip 1cm

\begin{spacing}{1.2}

\section{Introduction}\label{sec:intro}

The AdS/CFT correspondence \cite{Maldacena:1997re,Witten:1998qj}, first conjectured by Maldacena in 1997, provides a relation between gravitational theories with $(d+1)-$dimensional spacetimes that are asymptotically anti-de Sitter, and a $d-$dimensional non-gravitational gauge theory. This proposal has been one of the most important attempts to understand the interplay of quantum mechanics with gravity beyond the semi-classical approximation. One important program has been the development of the dictionary that describes how geometrical properties from the gravitational side get encoded in the information of the gauge theory. In principle, any physical quantity in the bulk theory should be obtainable from an observable in the boundary theory.

Perhaps the most remarkable element to consider is how the AdS/CFT correspondence changes our understanding of the nature of the metric. On the basis of this correspondence, many authors advocate for the emergence of the spacetime metric from fundamental quantum effects. For example, some approaches use the quantum entanglement in the dual field theory to recover information about the bulk metric \cite{Ryu:2006bv,Hubeny:2007xt,Dong:2016hjy,Balasubramanian:2013lsa,Headrick:2014eia,Czech:2014ppa,Czech:2015qta,Balasubramanian:2013rqa,Czech:2014wka,Myers:2014jia}, where the entanglement entropy is the dual quantity to the area of extremal surfaces in the bulk. To reconstruct the metric, these techniques compare the entanglement entropy of neighboring regions in the boundary. On the other hand, some other attempts \cite{Hamilton:2006az,Kabat:2011rz} make use of the equations of motion in the bulk. 

A new approach for reconstructing the bulk geometry from information in the boundary was recently introduced \cite{Engelhardt:2016wgb}. It was shown how the conformal metric of a holographic spacetime can be recovered from a set of distinguished spatial slices of the conformal boundary, known as ``light-cone cuts.'' Using these light-cone cuts, the conformal metric is reconstructed point by point in the bulk. The prescription shows how the metric can be recovered purely from the location of the light-cone cuts in the boundary. These light-cone cuts can be obtained from the dual field theory by analyzing the divergent behavior of $n$-point functions in the boundary, following a prior work in \cite{Gary:2009ae,Heemskerk:2009pn,Maldacena:1509iua}. Indeed, the $n$-point boundary correlation function has a divergence when all the points are null related to a single bulk vertex point, and energy-momentum at the vertex is conserved. This property can be used to determine the location of the vertex. This technique can be applied to almost\footnote{For static black holes, bulk vertices inside the photon sphere will in general fail to conserve energy-momentum from null related points at the boundary.} any point inside the causal wedge of the conformal boundary. 

It is of interest to inquire how the information gets encoded in the geometry of these slices in the boundary, apart from their location. Intuitively speaking, the light-cone cut of a bulk point $p$ is given by the collection of points $q$ in the boundary that are strictly null related to $p$, i.e., points $q$ for which there exists a null curve connecting $p$ to $q$ but not a timelike curve. In general, matter in the spacetime creates a gravitational lensing effect on the null light rays. If two null geodesics intersect, they form a caustic in the bulk. A point $q$ at the intersection of the boundary with any null geodesic that crosses a caustic inside the spacetime will fail to be strictly null related to $p$. The presence of these caustics creates a cusp in the light-cone cut at the boundary. This cusp is a distinguished conformal feature of the light-cone cut that encodes properties of the matter content of the bulk. 

We want to understand in detail how the matter content is encoded in the cusp. The simplest case to provide an example for this method consists of a spherically symmetric and static spacetime. We consider the instructive case of a superposition of a radiation star and a boson star, in asymptotic $AdS_4$. For one radiation star, the geometry is parameterized by its central density $\rho_c$, which determines the total mass of the spacetime $M_{Total}$. This mass can be recovered from an integral at infinity, allowing for $\rho_c$ to be recovered. To study a non-trivial example, we add a complex scalar field in the spacetime, creating a radiation/boson star in AdS. As shown in \cite{Astefanesei:2003qy}, boson stars in AdS are parameterized by the central value of the scalar field $\phi_c$, which is also correlated to the total mass measured in the boundary. By considering the superposition of both of them, the total mass is now parameterized by $\rho_c$ and $\phi_c$. In this case, knowing the total mass from the boundary only provides a relation between both parameters; however, it does not allow one to completely identify the spacetime. We will use the angle of the cusp to find a second relation between the parameters, which allows the identification of the matter content of the spacetime. 

The paper is organized as follows: in Section \ref{sec:Lightcut} we give a brief review of the light-cone cuts formalism, followed by the definition of a hyperbolic angle in the presence of a cusp on the light-cone cut. Next, Section \ref{sec:Star} discusses the model for a 4-dimensional boson star in AdS superposed with a gas of radiation. We provide numerical results that show how this geometry can be reconstructed from information on the boundary.  We conclude in Section \ref{sec:Discussion}.

\section{Light-cone Cuts} \label{sec:Lightcut}

\subsection{Review of the formalism}

This section provides a review of the pertinent properties of the light-cone cut formalism. We will follow the construction of \cite{Engelhardt:2016wgb}.

\begin{figure}
    \centering
    \begin{minipage}{0.45\textwidth}
        \centering
        \includegraphics[width=0.8\textwidth]{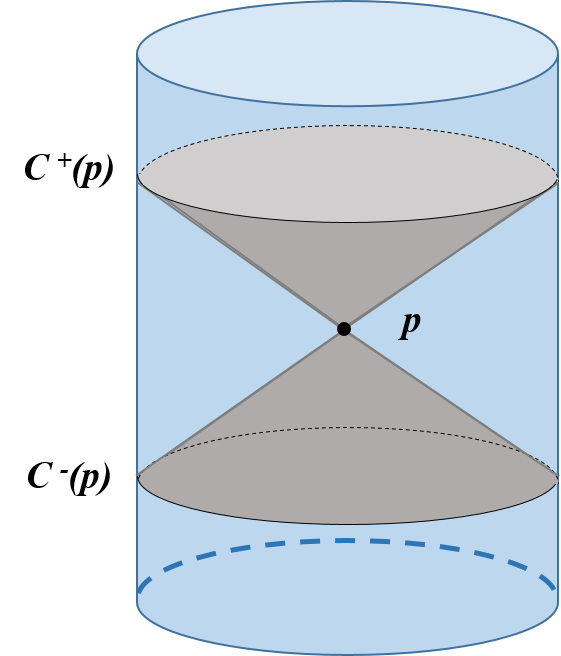} 
        \caption*{(a)}
    \end{minipage}\hfill
    \begin{minipage}{0.45\textwidth}
        \centering
        \includegraphics[width=0.87\textwidth]{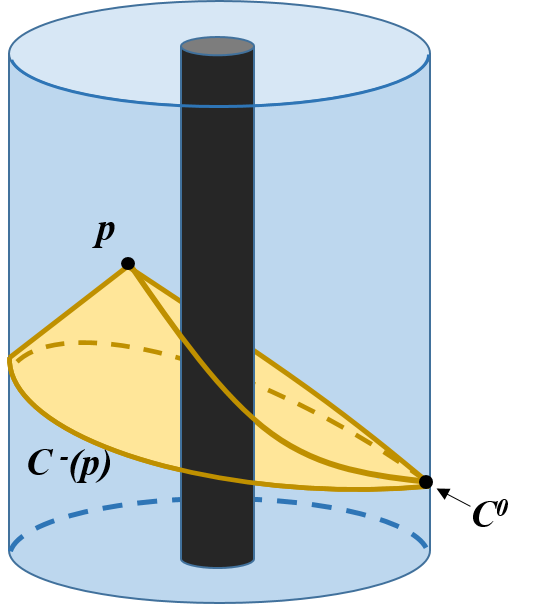}
        \caption*{(b)}
    \end{minipage}
    \caption{(a) Past and future light-cone cuts for a point $p$ in pure anti-de Sitter geometry. (b) With matter content or a black hole in the bulk, the past light-cone cut $C^-(p)$ is continuous, but fails to be smooth in the cusp created by gravitational lensing.}
  \label{fig:LightConeCuts}
\end{figure}

The chronological future of a point $p$, $I^+(p)$, is defined as the collection of all points that are connected to $p$ by a past-directed timelike curve. To include null curves, the causal future $J^+(p)$ of $p$ is defined to be the set of all points connected to $p$ through a past-directed timelike or null curve. Likewise, the chronological and causal past of a point $p$, $I^-(p)$ and $J^-(p)$ respectively, can be defined. We denote by $M$ an asymptotically locally AdS spacetime\footnote{We adopt the definition of asymptotically locally AdS spacetime from \cite{Fischetti:2012rd}. } with conformal boundary $\partial M$. In this construction, the spacetime $M$ is assumed to be $C^2$, connected, maximally extended, and AdS hyperbolic. In this work, we adopt the definition of AdS hyperbolic from \cite{Wall:2012uf}. This requirement corresponds to the absence of closed causal curves, together with the following condition: for any two points in the bulk $\{p,q\}$, if the AdS boundary has been conformally compactified, then the set $J^+(p) \cap J^-(q)$ is compact. The ``future light-cone cut" of a point $p\in M$, denoted by $C^+(p)$, is defined as the intersection of the boundary of the causal future of $p$ with $\partial M$: $C^+(p)\equiv \partial J^+(p) \cap \partial M$. The past light-cone cut of a point $p$, $C^-(p)$, can be obtained analogously. Figure \ref{fig:LightConeCuts} illustrates both the future and past light-cone cuts for a point inside pure AdS spacetime.

Some relevant properties of the light-cone cuts proved in \cite{Engelhardt:2016wgb} include:

\begin{itemize}
\item $C^{\pm}(p)$ is a complete spatial slice of $\partial M$.
\item $C^{\pm}(p)$ is a continuous set.
\item $C^{\pm}(p)$ is $C^1$ everywhere except on at most a measure zero set.
\item $C^{\pm}(p)$ correspond to a unique bulk point: $C^{\pm}(p)$ and $C^{\pm}(q)$ agree on an open set if and only if $C^{\pm}(p)=C^{\pm}(q)$.

\end{itemize}

It is important to notice that, when gravitational lensing is present, the light-cone cut is not given by the intersection of the entire light-cone with the boundary. The future light-cone of a point $p$ is defined as the set of all future-directed null geodesics fired from $p$. For a generic spacetime, null geodesics can focus and intersect, creating caustics. When a caustic is produced, the null geodesics leave the boundary of $J^+(p)$, and their intersection with $\partial M$ do not belong to the future light-cone cut of the point $p$. This property of caustics in general geometries implies that the light-cone cuts are not necessarily smooth slices of $\partial M$. Moreover, even though the light-cone cuts are continuous, the presence of cusps in the cut will yield points where it fails to be $C^1$. See (b) in Figure \ref{fig:LightConeCuts} for an illustration.

\subsection{Cusp angle}

The presence of cusps in light-cone cuts is a feature of general spacetimes due to the effect of gravitational lensing on the light rays. The details of these cusps are determined by the distribution of matter in the bulk. Therefore, features of the matter content inside the spacetime are expected to be encoded in the structure of the cusp at the boundary. To parameterize the cut, we define a deficit angle for the cusp in the light-cone cut. This angle will represent a conformally invariant attribute of the light-cone cut that can be measured at the boundary. 

In four dimensional geometries, the light-cone corresponds to a 3-dimensional Lorentzian hypersurface. For spherically symmetric spacetimes, using coordinates $\{t,r,\theta,\varphi\}$, we can further restrict the analysis to the equatorial plane $\theta=\pi/2$, reducing the light-cone to only two dimensions. We now introduce a new radial coordinate $\rho\equiv\arctan(r/L)$, where $L$ is the AdS radius, locating the boundary at $\rho=\pi/2$. After a conformal rescaling, the metric in the boundary is chosen to be the Einstein Static Universe, that is, the static cylinder $\mathbb{R}\times S^2$. Therefore, for $d=4$, the conformal boundary metric restricted to the equatorial plane is given by

\be \label{eq:BoundaryMetric}
ds^2_{\partial M} = -dt^2 +d\varphi^2
\ee
For this Lorentzian space, we can naturally define a hyperbolic angle. Let $u$ and $v$ be two spacelike vectors living in the boundary. We define the hyperbolic angle $\beta$ between them by:

\be \label{eq:Angle}
\cosh\beta=\frac{|u\cdot v|}{\sqrt{|u\cdot u||v\cdot v|}}
\ee
This angle corresponds to the absolute value of the boost parameter of the Lorentz transformation that relates both vectors, taking values in the interval $\left[ 0,\infty \right)$. The velocity of the associated boost is given by $v=\pm \tanh \beta$.

\section{AdS Radiation/Boson Star} \label{sec:Star}

The set of light-cone cuts enable us to recover the conformal metric of the bulk spacetime. We now study how properties of this geometry get encoded in the structure of the light-cone cuts at the boundary, specifically in the hyperbolic angle of the cusp, and how to use that information to reconstruct the bulk metric. 

According to the AdS/CFT dictionary, each bulk field has an associated dual operator in the boundary theory. Given a state in the CFT, for each non-vanishing one-point function of an operator in the boundary there exists a corresponding field in the bulk theory. In addition, when these one-point functions are static and spherically symmetric, their respective bulk fields will share these symmetries.

The simplest example of a static and spherically symmetric geometry that includes matter is a perfect fluid with a radiation equation of state. This acts as an idealized model for a star in AdS. For these geometries, an examination of the non-vanishing correlation functions of the dual operators in the boundary will exhibit the presence of a massless field in the bulk. In our approximation, we will consider a state of radiation of that field, modeled by a perfect fluid. These solutions form a one-parameter family, characterized by their central density $\rho_{c}$. One characteristic of these solutions is that the total mass is a function of the central density, $M_{Total}=M_{Total}(\rho_{c})$. The mass of the star can be computed as an integral at infinity, and thus the central density can be recovered from information on the conformal boundary. In order to provide a non-trivial example, where the cusp angle provides additional information, we will study the superposition of this radiation star with a boson star. 

Boson stars with a negative cosmological constant have been studied for many years \cite{Buchel:2013uba,Astefanesei:2003qy}. In particular, the spherically symmetric solutions form a family parameterized by the central value of the scalar field $\phi_c$. Similar to the case of a radiation star, the total mass of the geometry is a function of $\phi_c$. By adding a complex scalar field to the spacetime, we introduce a degeneracy in solutions with the same total mass but different $\rho_c$ and $\phi_c$. Computing the mass at the boundary provides a relation between the 2 parameters. The cusp angle provides a second relation between these parameters, and thus the spacetime will be uniquely identified.

\subsection{Model} \label{ssec:Model}

The action of an AdS radiation/boson star in four dimensions is given by

\be \label{eq:Action}
S=\frac{1}{8\pi G_{4}}\int_{M_{4}}d^{4}x\left(\sqrt{-g}\frac{1}{2}(R-2\Lambda)+\mathcal{L}_{boson}+\mathcal{L}_{rad}\right)
\ee
where the cosmological constant is given by $\Lambda=-3/L^{2}$, with $L$ the AdS radius\footnote{In this paper we use $c=L=1$.}.  This action describes the evolution of a complex scalar field $\Phi$ minimally coupled to gravity, superposed with a gas of radiation. The Lagrangian density for a complex scalar field is defined by

\be \label{eq:ScalarLagrangian}
\mathcal{L}_{boson}=\sqrt{-g}\left(g^{\mu \nu}\Phi_{,\mu}^{*}\Phi_{,\nu}+V(\Phi)\right)
\ee 
where $V(\Phi)$ is the potential for the scalar field, and $\Phi^{*}$ denotes the complex conjugate of the field. We will restrict ourselves to a potential without any self-interaction terms. For our case $V(\Phi)=\mu^{2}\Phi^{*}\Phi$, where $\mu$ is the mass of the scalar field. The Lagrangian density (\ref{eq:ScalarLagrangian}) is invariant under the transformation of the scalar field by a global phase, $\Phi\to\Phi e^{-i\alpha}$; this symmetry gives rise to a conserved current

\be \label{eq:Current}
J^{\mu}=ig^{\mu\nu}\left(\Phi_{,\nu}^{*}\Phi-\Phi_{,\nu}\Phi^{*}\right)
\ee
The corresponding conserved charge can be identified as the number of scalar particles,
 
\be \label{eq:ScalarParticles}
 N=\int d^{3}x\sqrt{-g}J^{t}
\ee 
Varying the action (\ref{eq:Action}) with respect to the fields $g_{\mu\nu}$ and $\Phi$ yields the field equations 
  
\begin{eqnarray} 
 R_{\mu\nu}-\frac{1}{2}g_{\mu\nu}R-3g_{\mu\nu} &	= &	8\pi G_{4}\left[T_{\mu\nu}^{rad}+T_{\mu\nu}^{boson}\right],\label{eq:EinsteinEquation}\\
\square^{2}\Phi-\mu^{2}\Phi &	= & 0, \label{eq:KleinGordon}
\end{eqnarray}
respectively. Here,

\be \label{StressRadiation}
T_{\mu\nu}^{rad}=\rho(r)u_{\mu}u_{\nu}+p(r)\left(g_{\mu\nu}+u_{\mu}u_{\nu}\right)
\ee
is the stress-energy tensor for a perfect-fluid gas of radiation, where in four dimensions the pressure is related to the energy density by $p(r)=\rho(r)/3$; and

\be \label{eq:StressBoson}
T_{\mu\nu}^{boson}=\partial_{\mu}\Phi^{*}\partial_{\nu}\Phi+\partial_{\nu}\Phi^{*}\partial_{\mu}\Phi-g_{\mu\nu}\left(\partial^{\alpha}\Phi^{*}\partial_{\alpha}\Phi+\mu^{2}|\Phi|^{2}\right)
\ee
is the stress-energy tensor for a boson star, where $\Phi=\phi_{1}+i\phi_{2}$ is a complex scalar field assumed to be harmonic $\Phi=\phi(r)e^{-i\omega t}$. In these solutions, both stress-energy tensors are spherically symmetric and time independent. It is advantageous to use the metric in the form

\be \label{eq:Metric} 
ds^{2}=-f(r)dt^{2}+h(r)dr^{2}+r^{2}d\Omega_{2}^{2}
\ee
where $f(r)$ and $h(r)$ are functions of the radial coordinate $r$, and $d\Omega_{2}^{2}$ is the line element of a 2-sphere. From the $tt$ component of equation (\ref{eq:EinsteinEquation}), using the ansatz $\Phi=\phi(r)e^{-i\omega t}$, and setting $8\pi G_{4}=1$, we obtain
  
\begin{multline}
 \frac{f(r)\left(h(r)\left[-1+h(r)+3r^{2}h(r)\right]+rh'(r)\right)}{r^{2}h^{2}(r)} =\\ f(r)\rho(r)+2\omega^{2}|\Phi|^{2}+f(r)\left(-\frac{\omega^{2}}{f(r)}|\Phi|^{2}+\frac{\left[\phi'(r)\right]^{2}}{h(r)}+\mu^{2}|\Phi|^{2}\right), \label{eq:ttComponent} 
\end{multline}
where the prime denotes derivative with respect to r. Similarly, from the $rr$ component of equation (\ref{eq:EinsteinEquation}), we obtain
 
\begin{multline}
 -\left(3+\frac{1}{r^{2}}\right)h(r)+\frac{1+rf'(r)/f(r)}{r^{2}}=\\ \frac{\rho(r)h(r)}{3}+2\left[\phi'(r)\right]^{2}-h(r)\left(-\frac{\omega^{2}}{f(r)}|\Phi|^{2}+\frac{\left[\phi'(r)\right]^{2}}{h(r)}+\mu^{2}|\Phi|^{2}\right), \label{eq:rrComponent}
\end{multline} 
 Additionally, the scalar field equation of motion (\ref{eq:KleinGordon}) yields
 
\be \label{eq:ScalarEOM} 
\frac{\omega^{2}}{f(r)}\phi(r)+\frac{1}{h(r)}\phi''(r)+\left[\frac{f'(r)}{2f(r)h(r)}-\frac{h'(r)}{2h^{2}(r)}+\frac{2}{rh(r)}\right]\phi'(r)-\mu^{2}\phi(r)=0
\ee
One further condition can be obtained from conservation of the stress-energy tensor. Assuming that the scalar field satisfies its equation of motion, the bosonic stress-energy tensor from equation (\ref{eq:StressBoson}) is necessarily conserved. Consequently, by requiring that the divergence of the total stress-energy tensor vanishes, an expression for the conservation of the radiation stress-energy tensor is obtained,
 
\be 
\nabla_{\mu}T_{rad}^{\mu\nu}	=	\nabla_{\mu}\left[\rho(r)u^{\mu}u^{\nu}+p(r)\left(g^{\mu\nu}+u^{\mu}u^{\nu}\right)\right]\\
	=	0
\label{eq:RadStressConservation}
\ee   
A useful expression is given by the radial component $\nu=1$:
 
\be \label{eq:RadStressConsCondition} 
 2\frac{f'(r)}{f(r)}=-\frac{\rho'(r)}{\rho(r)}
\ee
From equations (\ref{eq:ttComponent}), (\ref{eq:rrComponent}), (\ref{eq:ScalarEOM}), and (\ref{eq:RadStressConservation}), one can solve for the coefficients of the metric, $f(r)$ and $h(r)$, the radiation density $\rho(r)$, and the radial component of the scalar field $\phi(r)$. To solve these equations we need to impose appropriate boundary conditions. By requiring a regular, non-singular spacetime at the origin, the components of the metric must satisfy $f(0)=1$, and $h(0)=1$. Additionally, we demand the solution to be asymptotically $\text{AdS}_{4}$, giving the following boundary conditions at infinity:
 
\begin{eqnarray} 
 f(\infty)	&\to&	\left[1+r^{2}+\mathcal{O}(\frac{1}{r})\right],\label{eq:Asymptf}\\
h(\infty)	&\to&	\left[1+r^{2}+\mathcal{O}(\frac{1}{r})\right]^{-1},\label{eq:Asympth}
\end{eqnarray}
The boundary conditions for the scalar field are $\phi(0)=\phi_{c}$, and $\phi'(0)=0$, where $\phi_{c}$ is the central value of the scalar field. For the radiation energy density we have $\rho(0)=\rho_{c}$, with $\rho_{c}$ the value at the origin. For each value of the scalar field mass $\mu$, this family of solutions is described by two parameters: $\phi_{c}$ and $\rho_{c}$. Henceforth, we will set $\mu = 1$.
  
\begin{figure}
    \centering
    \includegraphics[width=0.8\textwidth]{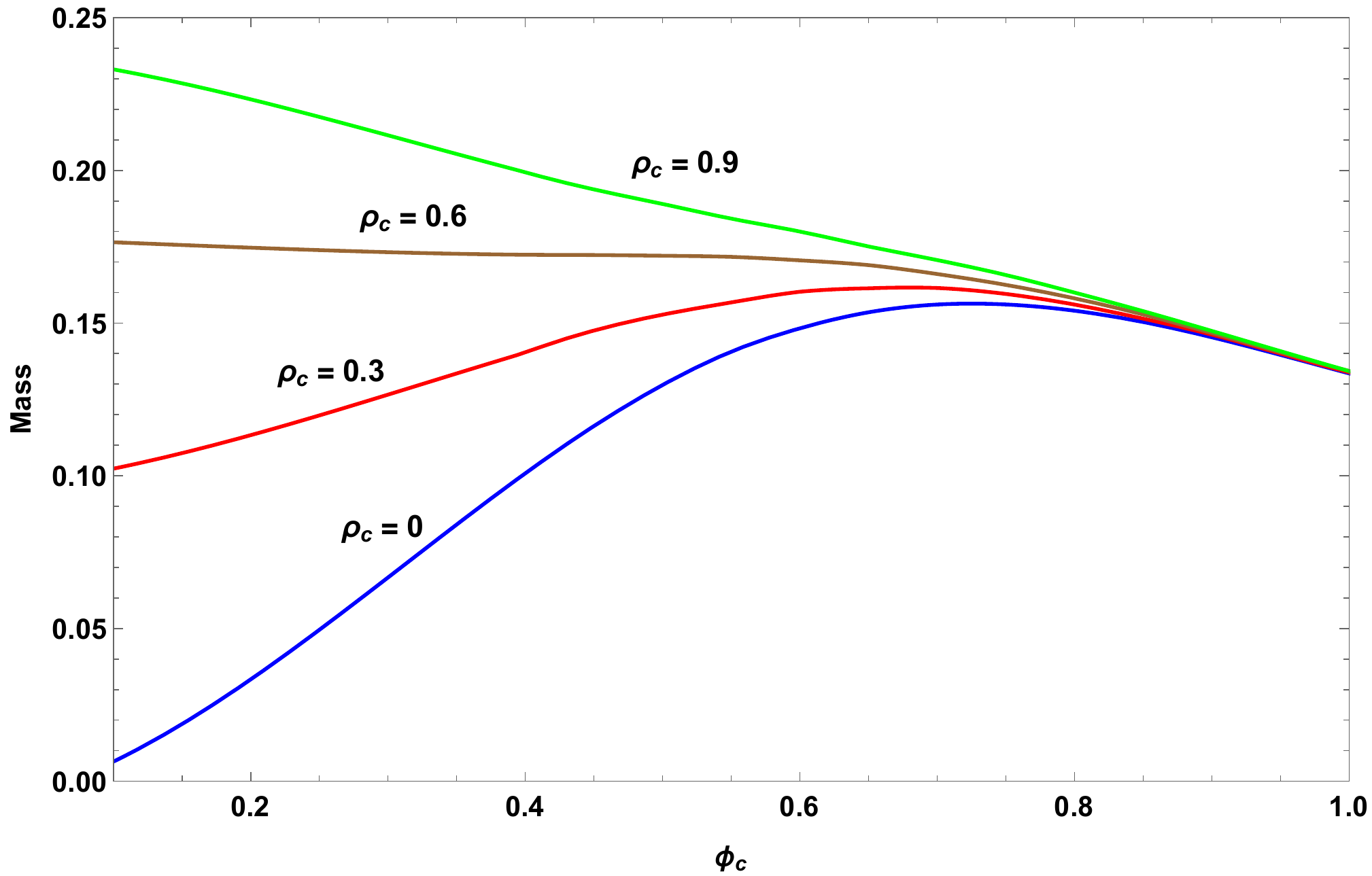}
    \caption{Mass of the spacetime as a function of $\phi_c$ and $\rho_c$.}
    \label{fig:MassPlot}
\end{figure}

Given a family of solutions, it is important to identify the configurations that yield a stable solution. Figure \ref{fig:MassPlot} shows the total mass $M_{Total}$ of the star as a function of $\phi_c$, for different values of $\rho_{c}$. In the pure boson star case $\rho_c =0$, the total mass is not a monotonic function of $\phi_c$. This suggests the onset of a dynamical instability. Studies have been done discussing the stability of boson stars in 4 dimensions with $\Lambda=0$,  \cite{Gleiser:1988rq,Gleiser:1988ih,Lee:1988av}. The transition between stability and instability of these solutions is located at the critical points in a plot of the mass as a function of the central value of the scalar field $\phi_c$. For boson stars in AdS, the stability of solutions was studied in \cite{Astefanesei:2003qy}, with similar results: the maximum mass $M_{max}$ represents the boundary between the regions of stability and instability. In essence, the transition from stable to unstable occurs when the solution admits a static perturbation (vanishing frequency $\omega=0$) that doesn't change the mass of the solution. Likewise, as noted in \cite{Hubeny:2006yu}, the self-gravitating gas of radiation has a similar behavior: for large central densities $\rho_c$, the mass becomes a non-monotonic function of $\rho_c$, suggesting the onset of a radial instability located at the maximum of the mass function. Following these results, in order to stay in a regime of stable configurations, only stars with $\rho_c\lesssim 0.6 $ and $\phi_c\lesssim 0.6$ will be considered.

\subsection{Geodesics}

The metric given by equation (\ref{eq:Metric}) has Killing fields associated to the time-translation symmetry and the azimuthal-rotation symmetry. In spherical coordinates $(t,r,\theta,\varphi)$, these correspond to $\xi=(1,0,0,0)$ and $\eta=(0,0,0,1)$, respectively. These Killing fields give rise to the energy $(e)$ and angular momentum $(l)$ as conserved quantities along any geodesic. 

Let $u^{\mu}=(\dot{t},\dot{r},\dot{\theta},\dot{\varphi})$ be the
four velocity of a given geodesic. Then, the normalization condition corresponds to  $g_{\mu\nu}u^{\mu}u^{\nu}=\kappa$, with $\kappa=-1$ for timelike geodesics, $\kappa=+1$ for spacelike geodesics, and $\kappa=0$ for null geodesics. In our case, as we are interested in null geodesics that reach out to infinity, the normalization condition in terms of the energy and angular momentum reads
\begin{equation}
f(r)h(r)\dot{r}^{2}=e^{2}-f(r)r^{2}\dot{\theta}^{2}-f(r)\frac{1}{r^{2}\sin^{2}\theta}l^{2}\label{eq:PseudoPotRadEq}
\end{equation}
Taking advantage of the spherical symmetry, it is convenient to restrict
ourselves to the equatorial plane $\theta=\pi/2$. In general, the geodesics will be parameterized by both $e$ and $l$. For null geodesics, it is possible to express these equations in terms of one parameter, which is $\alpha\equiv e/l$. By applying the chain rule, the geodesic equations can be written as functions of the radial coordinate: 

\begin{equation}
\frac{d\varphi}{dr}=\pm\frac{1}{r^{2}}\sqrt{\frac{f(r)h(r)}{\alpha^{2}-V(r)}},\ \ \ \ \ \ \ \ \ \ \ \ \ \ \ \ \frac{dt}{dr}=\pm\frac{\alpha}{f(r)}\sqrt{\frac{f(r)h(r)}{\alpha^{2}-V(r)}}.\label{eq:LightcutGeodesic}
\end{equation}
with $V(r)\equiv f(r)/r^{2}$. Then, in order to find the intersection of the light-cone with the boundary, it suffices to integrate these expressions out to infinity. In these equations, the sign depends on the radial component of the tangent vector at the initial point. For an ingoing radial component, we must use the negative sign; conversely, for an outgoing radial component, the positive sign is chosen.



\subsection{Recovering the metric}

Now we can analyze the process of recovering the geometry of our model of a radiation/boson star in AdS. First, the total mass $M_{Total}$ of the spacetime can be obtained through a boundary integral. Similar to the flat case, the asymptotic bulk metric can be expressed as $ds^2 = ds^2_0 + h_{\mu\nu}dx^{\mu}dx^{\nu}$ , where $ds^2_0$ describes the asymptotic AdS metric, and $h_{\mu\nu}$ represent the deviations from it. The mass and other conserved quantities are expected to be obtained from information encoded in $h_{\mu\nu}$. One approach for locally asymptotically $AdS_4$ geometries has been proposed in \cite{Aros:1999id}. The mass and the angular momentum can be obtained as Noether charges for the asymptotic Killing fields, $\partial / \partial t$ and $\partial / \partial \varphi$. Many authors have discussed different methods, including a Hamiltonian formalism, to compute the mass in asymptotically AdS spacetimes; for a further discussion, see \cite{Astefanesei:2003qy,Abbot:1981ff,Henneaux:1985tv,Chrusciel:2001qr,Ashtekar:1984zz,Aros:1999id,Aros:1999kt,Mielke:2001eh}. These different methods give equivalent results. By calculating the total mass of the spacetime $M_{Total}=M_{Total}(\rho_c,\phi_c)$, we obtain a relation between $\rho_c$ and $\phi_c$. 

The next step is to use $\beta$ from equation (\ref{eq:Angle}), the angle of the cusp in the light-cone cuts, to get a second relation and disambiguate for the exact value of both parameters. In this analysis, we use the cusp in the future light-cone cut $C^+(p)$. For fixed $\rho_c$ and $\phi_c$, the angle of the cusp can only depend on the initial position of the point $p$, given by the radial coordinate $r_0$. This quantity relies upon the choice of a coordinate system for the bulk. We can find a more invariant description of the point using the time coordinate on the boundary by considering the difference $\Delta t = t_{max} -t_{min}$, where $t_{min}$ and $t_{max}$ are the time coordinates of the first and last point to reach the boundary in the light-cone cut, respectively. For a point $p$ located in the middle of the spacetime, spherical symmetry guarantees a spherically symmetric light-cone cut. In this case, the time coordinate in the boundary is the same for every point in the light-cone cut; hence $\Delta t = 0$. As the point $p$ moves towards the boundary, the difference $\Delta t$ increases, and in the limit $r_0 \to \infty$ we recover the result associated to the light-cone cut being traced by a null geodesic traveling in the boundary, namely $\Delta t = \pi$. For the AdS radiation/boson star, the time difference is a monotonic function of the initial position $r_0$. Therefore, the position of the point $p$, and its respective light-cone cut, can be parameterized by $\Delta t$. 

Using the static time coordinate in the bulk $t_{bulk}$, there is a preferred rest frame in the dual theory for which the bulk Killing vector $\partial / \partial t_{bulk}$ has unit norm when it gets extended to the boundary. This provides a natural choice of time coordinate $t$ in the boundary theory.

\begin{figure}
    \centering
    \includegraphics[width=0.8\textwidth]{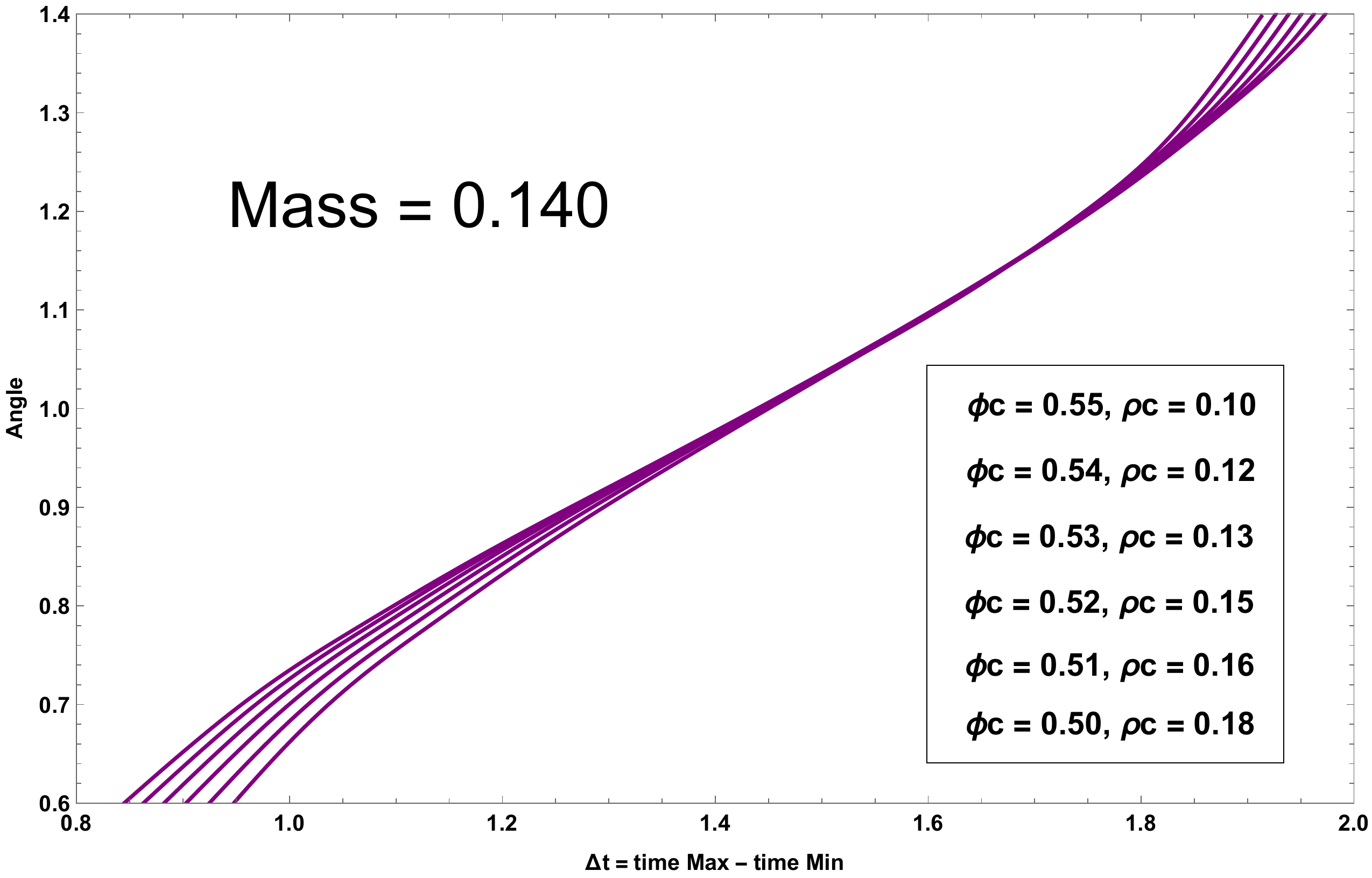}
    \caption{Plot of the angle $\beta$ vs $\Delta t$ for solutions with different values of $\rho_c$ and $\phi_c$, and total mass $M_{Total}=0.140$.}
    \label{fig:Results1}
\end{figure}

\begin{figure}
    \centering
    \includegraphics[width=0.8\textwidth]{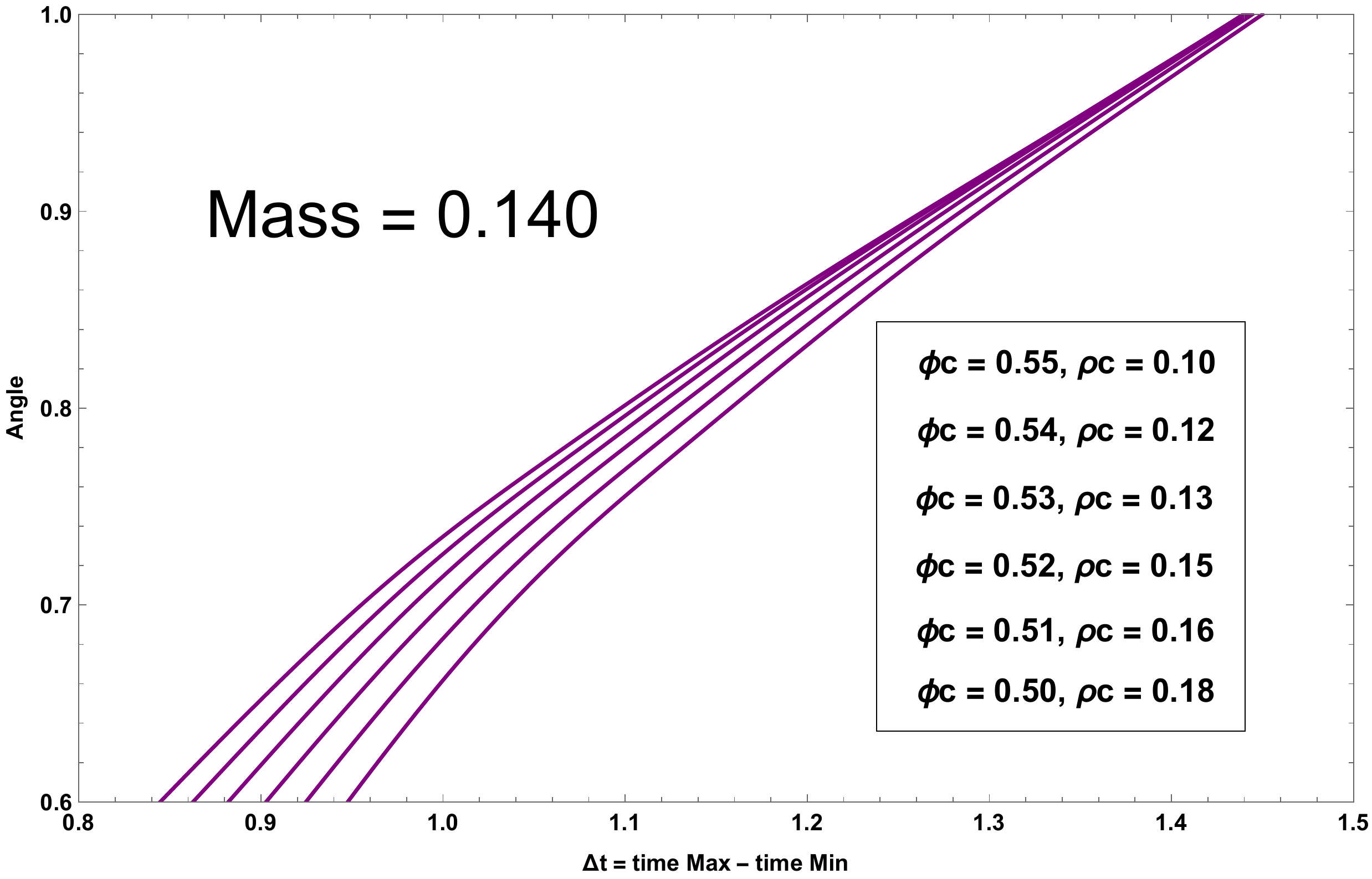}
    \caption{Close-up to the plot of the angle $\beta$ vs $\Delta t$ for solutions with different values of $\rho_c$ and $\phi_c$, and total mass $M_{Total}=0.140$.}
    \label{fig:Results2}
\end{figure}

\begin{figure}
    \centering
    \includegraphics[width=0.8\textwidth]{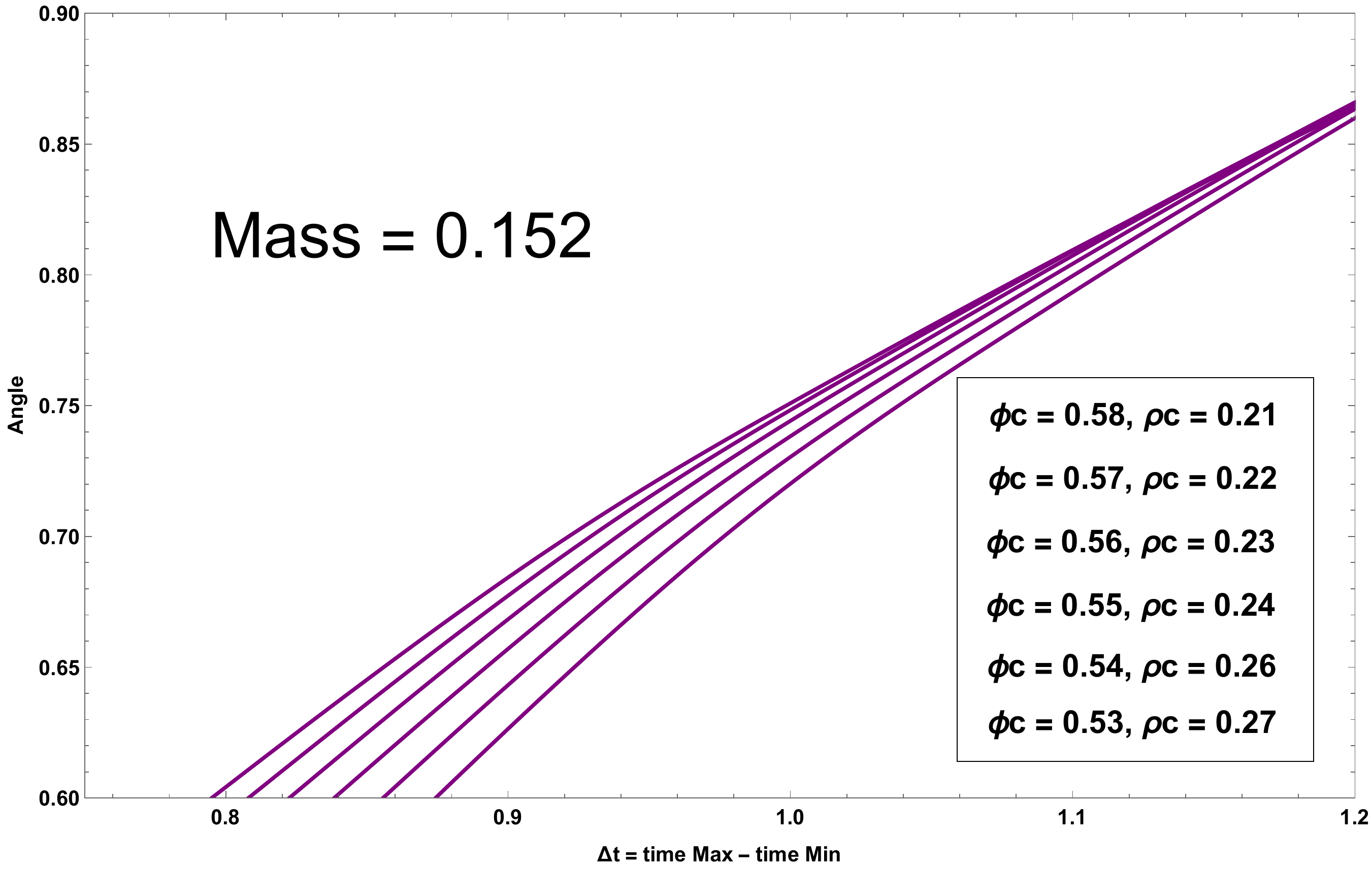}
    \caption{Close-up to the plot of the angle $\beta$ vs $\Delta t$ for solutions with different values of $\rho_c$ and $\phi_c$, and total mass $M_{Total}=0.152$.}
    \label{fig:Results3}
\end{figure}

\begin{figure}
    \centering
    \includegraphics[width=0.8\textwidth]{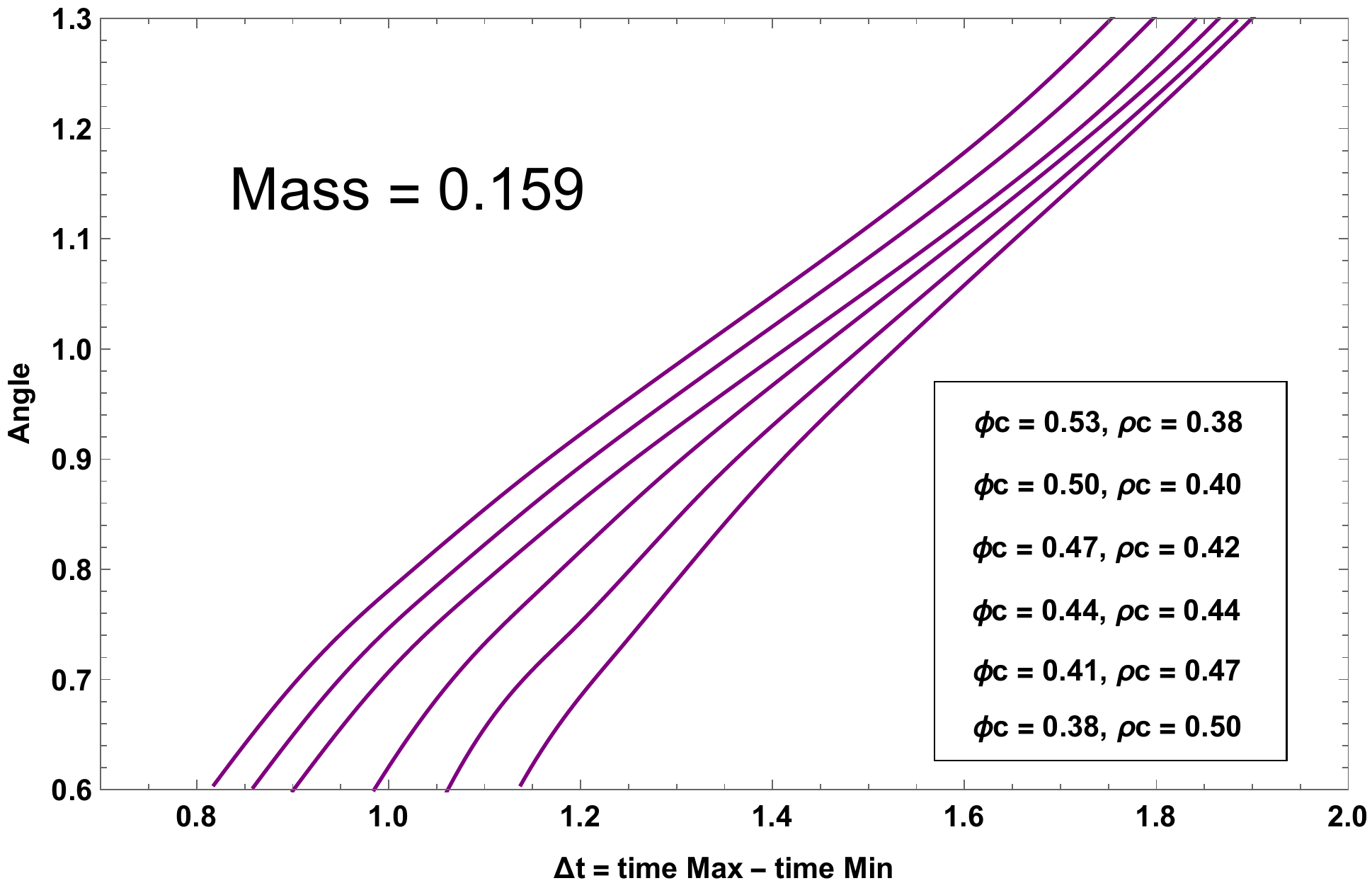}
    \caption{Close-up to the plot of the angle $\beta$ vs $\Delta t$ for solutions with different values of $\rho_c$ and $\phi_c$, and total mass $M_{Total}=0.159$.}
    \label{fig:Results4}
\end{figure}

For a fixed $M_{Total}=M_0$, different parameters $\rho_c$ and $\phi_c$ yield a different plot $\beta$ vs $\Delta t$. These plots for geometries with $M_0$ can be compared: if none of the plots coincide with each other, identifying the particular plot gives the specific values of the parameters. Figure \ref{fig:Results1} shows the plot of the cusp angle as a function of $\Delta t$, for different values of $\rho_c$ and $\phi_c$, all for the same total mass $M_{Total}=0.140$. As the radial coordinate of the bulk point approaches the origin, $\Delta t$ goes to zero, the cusp becomes flat, and the angle $\beta$ approaches zero. Conversely, as the radial coordinate of the bulk point approaches the boundary, $\Delta t$ reaches the maximum value of $\pi$ and the cusp angle becomes infinite. The latter case corresponds to the light-cone cut being created by light rays traveling in the boundary. The numerical results show that for the same mass, different values of $\rho_c$ and $\phi_c$ yield distinct curves. Figure \ref{fig:Results2} shows a close-up view of Figure \ref{fig:Results1}, in which different geometries with the same mass $M_{Total}=0.140$ are compared. As we can see in Figure \ref{fig:Results2}, there exists a region in the plots where each solution can be clearly identified: for a fixed value of $\Delta t$ in this region, each solution has a different angle $\beta$. Figure \ref{fig:Results3} and \ref{fig:Results4} show the same pattern for solutions with total mass of $M_{Total}=0.152$ and $M_{Total}=0.159$, respectively. As discussed at the end of section \ref{ssec:Model}, all the solutions considered in these figures have parameters $\phi_c$ and $\rho_c$ within the region of stability. These numerical results exhibit how the cusp angle can be used to recover a second relation between the two parameters. Thus, the matter content and the geometry of this non-trivial bulk spacetime can be reconstructed purely from boundary information.

\section{Discussion} \label{sec:Discussion}

We have presented a non-trivial example in which the geometry of the bulk can be reconstructed entirely from information in the boundary. We have used the light-cone cut construction of \cite{Engelhardt:2016wgb} and explored more features of the light-cone cuts. In the presence of matter, gravitational lensing will create caustics in the spacetime, generating cusps in the light-cone cuts. In this paper, we have explored and taken advantage of these cusps. By using the deficit angle of these cusps, we have included additional information from the boundary that allowed us to reconstruct the geometry of a superposition between a radiation and a boson star.

\begin{figure}
    \centering
    \begin{minipage}{0.45\textwidth}
        \centering
        \includegraphics[width=0.8\textwidth]{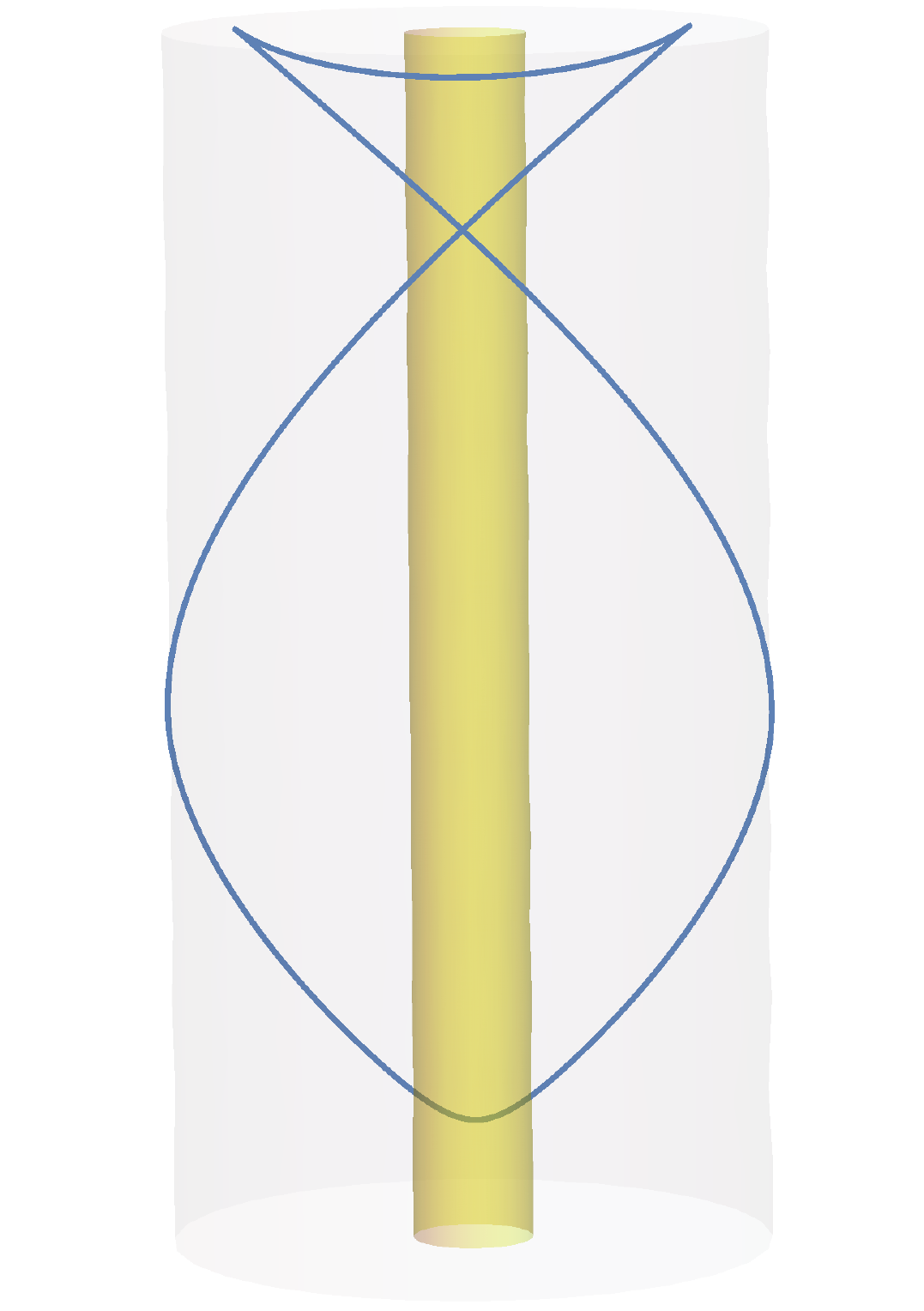} 
        \caption*{(a)}
    \end{minipage}\hfill
    \begin{minipage}{0.45\textwidth}
        \centering
        \includegraphics[width=0.8\textwidth]{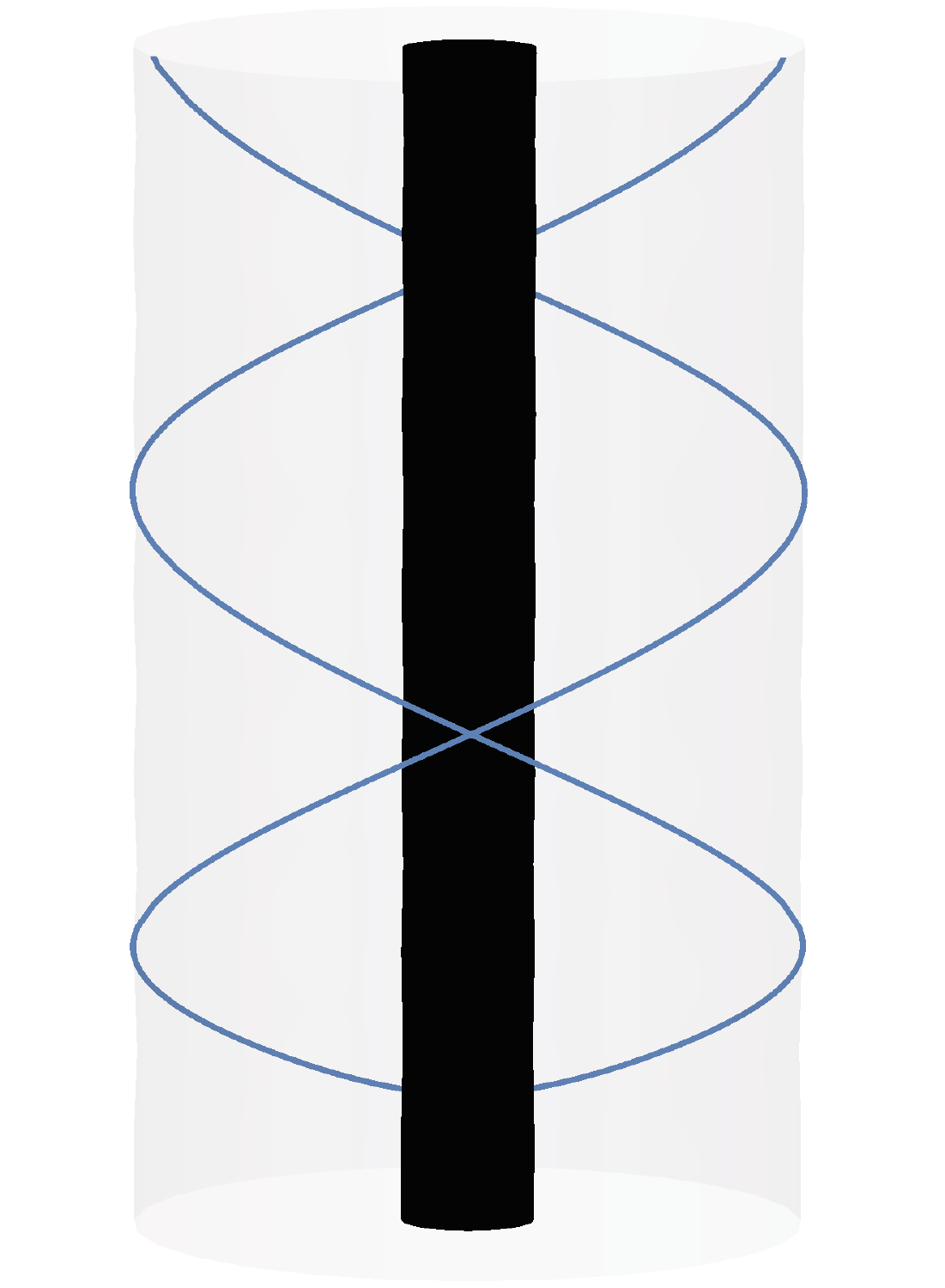}
        \caption*{(b)}
    \end{minipage}
    \caption{Intersection of the light-cone and $\partial M$ for: (a) Self-gravitating radiation gas in AdS, and (b) AdS Eternal Black Hole.}
  \label{fig:UltracompactObjects}
\end{figure}

The light-cone cuts of the AdS radiation/boson star have the structure of the image (a) in Figure \ref{fig:UltracompactObjects}. In this figure we plot the intersection of the complete light-cone with the boundary of the spacetime; we will refer to this subset of the boundary as the ``extended light-cone cut''. This image shows the existence of three angles: one in the cusp of the light-cone cut, and two more angles on the top of the extended light-cone cut, one on each side. In the latter case, one can use equations (\ref{eq:LightcutGeodesic}) to find the value of the coordinates $t$ and $\varphi$ of the null geodesic at the boundary. The locus of points $(t_{Bndy},\varphi_{Bndy})$ describes a curve in the boundary as a function of $\alpha$. As mentioned in \cite{Hubeny:2006yu}, the slope of this curve is given by $\(dt/d\varphi\)_{Bndy}=1/\alpha$. Indeed, in the case of an AdS radiation/boson star, we have obtained numerically this same result. This shows that the derivative of the curve on each side of the meeting point of the top angles arrives with the same value, giving a cusp with zero angle. Therefore, these additional angles don't contribute any additional information about the matter content of the spacetime.

Another feature of the bulk that can be identified from the light-rays structure is the existence of ultracompact objects. An ultracompact object is defined as one with a radius smaller than $3/2$ its Schwarzschild radius, and so it exhibits a photon sphere. Defined this way, black holes are ultracompact objects, but there are also other non-black holes objects that admit a photon sphere \cite{Iyer:1985}. The presence of a photon sphere leaves a characteristic imprint in the extended light-cone cut.  The intersection of the boundary $\partial M$ and the light-cone of a point $p$ located outside the photon sphere radius $R=3R_{Schw}/2$ is generated by all the null geodesics that remain outside of the photon sphere. The geodesics that pass close to this sphere will travel around the object several times before coming out to infinity. Given that they can travel arbitrarily close to this orbit, these geodesics experience a considerable time dilation, and so the time that it takes for one of these geodesics to come back out to infinity can be arbitrarily large. This effect leaves a signal in the form of an ongoing, extended light-cone cut, which intersects itself an infinite number of times. On the other hand, for non-ultracompact objects, the intersection of the light-cone with the boundary of the spacetime will be finite. Figure \ref{fig:UltracompactObjects} shows the extended light-cone cuts for two different asymptotically AdS geometries: (a) a self-gravitating radiation star, and (b) an eternal black hole. In the first case, the extended light-cone cut is finite; in the second case, this extended cut intersects itself an infinitely number of times, and it is unbounded in the time direction. Using this feature, and having access to the extended light-cone cut, one can identify ultracompact objects in the bulk from information in the dual theory at the boundary. A different approach to probe ultracompact objects in the bulk was studied in \cite{Hubeny:2006yu}. In this previous work, the authors discussed the relation between singularities of two-point functions and several causal properties of the bulk spacetime. In particular, they considered points in the boundary that are connected causally by a bulk null geodesic. With these ``bulk-cone singularities", they argued how to use them to probe the bulk geometry in the context of different asymptotically AdS spacetimes, including ultracompact objects. 

An avenue of interest for future work would be to consider how the light-cone cuts could be used to identify black holes from other ultracompact objects. 

\section*{Acknowledgements}

It is a pleasure to thank Gary Horowitz for very helpful discussions. We also thank Seth Koren for proofreading an earlier version of this manuscript, and Veronika Hubeny for a useful comment. This work was supported in part by NSF Grant PHY-1504541, and by UC MEXUS-CONACYT Doctoral Fellowship.

\end{spacing}

\bibliographystyle{JHEP}
\bibliography{all}

\end{document}